# Josephson junction decoupling is the main origin of AC resistivity in the superconducting state

S. Sarangi[*], S. P. Chockalingam, S. V. Bhat

Department of Physics, Indian Institute of Science, Bangalore-560012, India

[*]Corresponding author:

Subhasis Sarangi

Department of Physics

Indian Institute of Science

Bangalore – 560012, India

Tel.: +91-80-22932727, Fax: +91-80-23602602

E-mail: subhasis@physics.iisc.ernet.in

**Abstract:**

The origin of AC resistivity in the high $T_c$ superconductors is not addressed adequately in literature. We found out, Josephson Junction (JJ) decoupling is the main origin of the AC resistivity in high $T_c$ superconductors. We have measured the AC resistivity in the superconductors in the low frequency range by measuring the complex AC impedance of superconducting $YBa_2Cu_3O_7$ (YBCO) polycrystalline samples. Our data shows that under certain conditions when the number density of Josephson Junctions (JJ) present in the sample and the JJ critical current crosses a threshold value, AC resistivity in the superconducting state keeps on increasing with lowering temperature. The underlying mechanism is an interesting interplay of JJ coupling energy, amplitude of the supply AC voltage and the current applied to the superconducting sample. The effect of the applied AC current of different frequencies and the variation of temperature were studied in detail. To find out the exact relation between the JJ coupling energy, JJ number density, applied AC frequency, the amplitude of AC current and the AC resistivity in the superconductors, we have studied samples of different grain sizes, pressurized with different pressure and sintered at different physical situations. These results have important implications for the understanding of the origin of AC resistivity and characterization of superconducting samples. In this paper we also extend the capability of the AC impedance studies in superconductors for the characterization of materials for device applications.

**Introduction:**

Zero resistance in superconducting materials means that there is no voltage drop along the material when a current is passed through, and by consequence no power is dissipated. This is, however, strictly true only for a DC current of constant value. This is not true in case of AC. The exact reason for the appearance of resistance and the power dissipation in case of AC is not addressed properly in literatures. Passing AC in superconducting materials without dissipation is one of the major challenges for the various applications of superconductors. The application of a magnetic field and the variation of temperature alter the AC penetration depth of the superconducting sample, which in turn changes the AC impedance associated with it. We have studied the AC resistivity of various superconductors to find out its origin in details. We found out, JJ decoupling is one of the most important factors controlling the AC resistivity in superconducting samples. JJ decoupling is basically the breaking of a JJ to normal state either by applying current or magnetic field or heat energy. Energy is absorbed or emitted in the formation of annihilation or creation of the Josephson junctions in the prices of JJ decoupling.

The AC-conductivity or resistivity in isotropic disorder systems has been extensively studied [1, 2, 3, 4, 5, 6], several experiments support the existence of unique relation in a variety of materials, in either electronic [3, 7] or ionic [3, 4, 5, 6] or superconducting [8, 9, 10, 11, 12, 13, 14, 15, 16, 17, 18] systems. For both application and fundamental reasons, investigation of the frequency variation of the Ohmic and reactive components of the linear conductivity, $\sigma' + i\sigma''$, represent one of the important approaches to understand the nature of high $T_c$ superconductors. Ac-conductivity in superconducting system gives new phenomenon like non-Ohmic dissipative regime in YBCO [8], conducting coherence peak [9, 10] and the evidence of Bose-glass transition in superconductors [11, 12]. At low frequency, in most of the system AC resistivity is frequency independent. In general it is expected that the AC or the DC resistivity in superconductors should decrease below $T_c$. In DC resistivity case the value of resistance goes exactly to zero and in the case of AC resistivity it keeps on decreasing towards zero with decreasing temperature. But surprisingly we found in certain type of superconducting samples, the AC resistivity drops very close to zero at the critical temperature and instead of decreasing it increases slowly with decreasing temperature below the critical temperature, which property is also strongly dependent on the applied AC frequency. Investigation of the above phenomena gave us the information regarding the contribution of JJ decoupling towards the AC resistivity of superconducting samples.

Measurements of the dynamic responses in superconductors just below the superconducting phase transition have provided important information on the nature of the superconduting state. Complex AC conductivity has been measured in different polycrystalline YBCO sample at frequencies starting from 100 Hz to 100 kHz and temperature 10 K to 300 K. Here we explain the results with the decoupling of Josephson junctions present in the superconducting samples. Transport-critical-current behavior and non-Ohmic dissipative regime in superconducting YBCO is strongly affected by this JJ decoupling [13, 14]. Josephson junctions are intrinsically present in all types of superconducting sample. It is found out that even single crystals made with almost all cares are not free from intrinsic JJs or weak links. The number density of JJ depends on the sample type and the sample quality. It also depends on the stoichiometry of sample preparation. The number density of JJ present in granular polycrystalline samples is more than thin film and single crystalline samples. Contacts between the grains in polycrystalline samples and the defects, scratches and the layer structures in the single crystalline samples are the main sources of Josephson junctions. The influence of the Josephson junction decoupling on the complex conductivity is studied in detail.

The electromagnetic absorption or the dissipation of any means inside the superconductors belongs to the same class of behavior as the AC resistivity. Electromagnetic absorption or the dissipation results from currents that are in phase with the electric field and hence is proportional to $Z'$, the real part of AC resistivity. So the electromagnetic absorption can be expected to occur in $Z'$ for all the frequencies well below the energy gap. JJ decoupling is one type of loss like the electromagnetic absorption, which can also be represented as the AC resistance in the system. In the present work, we have measured the AC loss due to the JJ decoupling and converted it to the equivalent AC resistance for the system. In this way we have correlated the JJ decoupling with the increase of the AC resistivity below $T_c$ in certain superconductors.

**Experiment:**

In our experiments we investigated in the high $T_c$ cuprate superconductor YBCO. The YBCO powder used was made by solid-state reaction of reagent-grade $Y_2O_3$, $BaCO_3$ and CuO. We prepared two pellets made of polycrystalline samples of YBCO. Both the pellets were exactly similar in size and shape but prepared with different techniques. One of the two pellets (Sample 1) is perfectly sintered and all the precaution has been taken care to minimize the presence of Josephson junctions in it whereas the other sample (Sample 2) is prepared keeping in mind to make it granular in nature and to enhance the possibilities of Josephson junctions. Both the pellets were prepared from the same phase of YBCO powder. First pellet was prepared from the YBCO powder after grinding further for 5 hours and pressurizing with a pressure of 2 tons. The second pellet was prepared from the same YBCO powder without grinding further but sintered at temperature of 980 $^0C$ in flowing Oxygen for 8 hours. Both the samples show sharp superconducting transition temperature at ~91 K as determined by $\rho$~$T$ and ac susceptibility measurements. The materials were found to be single phasic as determined by x-ray diffraction. The SEM image of both the samples surfaces is shown in Fig. 1. The SEM image of sample 1 shows a much more dense YBCO microstructure than sample 2. This dense microstructure in the sample 1 reduces the number density of JJs present in the sample and also makes the existing JJs stronger. The granular nature in sample 2 is more favorable for weak JJs. So in the sample 2 the number density of JJ is more than sample 1.

Four terminal measurements were performed with a standard dual phase lock-in-amplifier (model SR830) with two-phase detections. The current and voltage leads were soldered onto the surface of the samples with silver paint contacts. The contact resistances were estimated to be below 2 Ω. The AC current was monitored across a standard resistor in series with the sample, and the sample voltage was measured using the lock-in-amplifier.

The sample was arranged inside an Oxford instrument cryostat. No precautions were taken to expel the earth magnetic field. The AC resistivity measurement was performed on both the sintered and non-sintered pellets (sample 1 and sample 2) having equal size. Rectangular bars with dimensions of 11 × 4 × 3 mm were cut from the sample pellets for the AC resistivity measurements. Extensive impedance studies have been performed from 10 to 300 K. The resistivity was measured in the frequency range 100 Hz < $f$ < 100 KHz. We carefully checked the dispersive behavior of the leads using some standard cells (metallic samples with very low resistance) and ensured that there was no extraneous inductive or capacitive coupling in these frequencies ranges. Measurements were made after stabilizing the temperature for about 10 min prior to each reading.

**Results & Discussion:**

Figure 2 shows the real part of the AC resistivity of both the samples (sample 1 and sample 2) measured at frequency of 100 KHz. The superconducting transitions are clearly visible in both the samples at the temperature of 91 K. Just below transition temperature sample 1 shows AC resistivity nearly zero upto the temperature of 13 K, but the sample 2 shows AC resistivity of 6.5 μΩ cm at the temperature of 13 K and follows a Ambegaokar-Baratoff function for tunnel junction [19] like behavior below $T_c$. Inset of figure 2 shows the clear picture of the AC conductivity below $T_c$ upto 13 K. The polynomial fitting in the inset of figure 1 confirms that the plot nearly follows the Ambegaokar-Baratoff function. At normal state DC resistivity behavior for both the samples follow the patterns of a standard YBCO sample. Sample 2 shows higher resistivity than sample 1, this may be due to the granular nature of the sample 2. To know details about the frequency dependence of the real part of AC resistivity of the sample 2, we have done experiments at various frequencies in between 100 Hz to 100 kHz. The AC resistivity below $T_c$ for different frequencies is presented in Fig. 3. Here it is found that at all the higher frequencies, it follows nearly Ambegaokar-Baratoff function similar like the inset of Fig. 1. The amplitude of the dependence increases with frequency. At lower frequencies plots and also at the temperature just below $T_c$ for all the plots, the behavior does not follow Ambeogkar-Baratoff function properly. This is either due to the low intensity of the signal buried by the background noise or the difference of the JJ tunnel current behavior in high $T_c$ superconductors with the Ambegaokar-Baratoff function. This is discussed in details latter. The behavior of the in-phase part of the resistivity (real component) with frequency for both the samples is shown in Fig. 4 at the temperature of 30 K. The in-phase part of the resistivity of the granular sample (sample 2) increases with frequency in the range of frequency from 100Hz to 100 KHz with an average change of 55 X $10^{-12}$ Ω *cm* / Hz, but for the sintered sample (sample 1) the resistivity increases with an average change of only 4 X $10^{-12}$ Ω *cm* / Hz. This is very less comparatively to the sample 1. But in both the cases the fitting is linear which shows the deep interconnection between the real part of the resistivity and the frequency of applied AC current.

It is important to understand the differences in the real part of the conductivity between both the samples and their frequency dependences; we have given the following explanation. Between the two superconducting samples, sample 1 is the standard superconducting sample, which shows the real part of the AC resistivity remains zero throughout the superconducting state (Fig. 2). As it is already discussed that sample 2 is a specially prepared sample to enhance the number density of Josephson junctions and the grain boundary weak links. Sample 2 can be model as an array of weakly Josephson-coupled, strongly superconducting anisotropic grains. The intergranular weak links are not only insulating barriers but instead proximity junctions, coupled via semiconducting, normal conducting, or poorly superconducting materials. Due to the occurrence of the large number of weak Josephson junctions in the sample 2, JJ decoupling in the sample 2 is more and due to this, it shows the real part of AC resistivity increases with lowering temperature in the superconducting state (Fig. 2). To know more details lets explain how these JJ decoupling gives these results; first let us consider what will happen to a single Josephson junction when we pass current more than its critical value. The obvious answer is it will break the junction and the junction will become normal. The AC current we apply in the superconductor mostly flows in the surface of the superconducting sample depending on the penetration depth λ. So the AC current flow through out the surface of the sample including the JJs present in the surface. Here we assume that the amount of AC current flowing in the sample wont be able to break all the JJs but only breaks the weaker JJs which have critical current less than the applied AC current. As earlier discussed each Josephson junction is associated with some coupling energy. This is known as JJ decoupling energy. JJ decoupling energy $E_j$ between neighboring grains is given by [14]

$$E_J(T,H) = \frac{\hbar}{2e} F(T) \langle I_0 \left| \frac{\sin \pi \phi / \phi_0}{\pi \phi_0} \right| \rangle \quad \text{------- (Eq. 1)}$$

Where $\Phi_0 = ch/2e$ is the flux quantum and $F(T)$ is a function of the temperature which in the Ambegaokar-Baratoff theory [19] is given by...

$F(T) = \{\Delta(T)/\Delta(0)\} \tanh \{\Delta(T)/2K_BT\}$ ------------------- (Eq. 2)

with $\Delta(T)$ the temperature dependent gap parameter and $\Delta(0)$ the gap at $T = 0$; $I_0$ is the maximum Josephson current given by

$$I_0 = \frac{\pi \Delta(0)}{eR_n}$$ ---------------- (Eq. 3)

with $R_n$ the normal state resistance of the junction; $\Phi = HA_J$ and $A_J$ is the effective field penetration junction area orthogonal to $H$. In the average Eq. (1) refers to the statistical distribution of the junction geometrical parameters. So once the Josephson junction breaks, the system will absorb energy $E_j$, this is a function of both temperature and the magnetic field. To make the experiment simple, we have eliminated the magnetic field $H$ contribution by keeping the field very near to zero. The exactly opposite situation will happen when we reduce the current across the critical current. The JJ, which was decoupled before by the application of higher current, will come back to its original situation and the system will release exactly same amount of energy $E_j$ just below $J_c$ and will become a Josephson junction again. If we pass an AC current in the Josephson junction having amplitude more than the critical current of Josephson junction then the system will absorb and release energy continuously as represented in the figure 5. At the point 'a' the current passes through the junction is more than its critical value, so the junction will absorb energy and at the point 'b' it will emit the energy in the form of heat. So the expression for the total energy absorbed by the sample per second or the total loss can be expressed as $P$ and $P = 2fNE_j$ (Eq. 4), where $f$ is the frequency of the AC current,

*N* is the total number of Josephson junctions keeps on breaking and forming due to the application of AC current and $E_j$ is the JJ decoupling energy of a single JJ.

The total energy absorbed by the sample due to the passing of AC current is a measure of loss and can be converted as the resistance of the system. So using the simple relation $P = I^2R$, where *P* is the total power, *I* is the applied current and *R* is the resistance of the sample, we can get the equivalent AC resistance $R_{ac} = P/I^2 = (2fNE_j/I^2)$ -----(Eq. 5). So $R_{ac}$ is proportional to the excess loss and, consequently, to the inverse quality factor $Q-1$. From this explanation (see Eq. 5) it is understood that if the sample have more JJs or higher the JJ critical current (not more than the applied AC current because in that case the JJ does not decouple) then the loss will be more and consequently the AC resistance will be more. From the Fig. 1 it is clear that the sample 2 has more number of JJs than the sample 1, so it shows AC resistivity more and due to the presence of very less number of JJs in sample 1 it shows the AC resistivity very near to zero at the same range of frequency. Now to explain the Ambeogkar-Baratoff (AB) pattern of the AC resistivity in the sample 2, we have to look at the equation 2. From the equation 2 it is clear that the critical current of a JJ and the $E_j$ both follows the AB pattern. So the AC resistivity, which is a linear function of the JJ decoupling energy $E_j$, has to follow the AB pattern. Just to explain the temperature dependent AC resistivity (real part) of the granular YBCO sample (sample 2) at various frequencies (Fig. 4), we have to see the equation 5 again. From the equation 5 it is clear that the AC resistivity in the superconducting sample is directly proportional to the frequency of applied AC current. Increasing frequency makes the total number of JJ decoupling per second more so the energy loss or the AC resistivity due to the JJ decoupling increases linearly with the applied frequency.

Here in the above discussion we have assumed all the JJs to be have equal critical current but in the real picture, there will be a distribution of JJ critical current for the whole JJs and that may be one of the reasons why the experimental curves discussed above do not fit exactly with the Ambeogkar-Baratoff function. The Ambeogkar-Baratoff function is applicable for a single Josepson junction. It is possible to give a model considering the whole JJs network in the sample and its corresponding fit and the fit may exactly fit with the experimental data. But the work is in progress and will be discussed in a forthcoming publication [20].

**Conclusions:**

We have intentionally varied microstructures of polycrystalline YBCO samples. We have observed close correlation between AC conductivity and the microstructures of the superconducting sample. In the view of the discussion made in the previous section it is concluded that the result of the study of the transport properties of the superconductor YBCO by AC impedance are consistent. In conclusion, we have observed a well-defined increase in the real part of the finite-frequency AC resistivity in the granular superconductor. The AC response obtained below critical temperature in the granular superconducting sample is mainly due to the JJ decoupling. Which of this effect is responsible for the difference between the resistivity and relaxation time remains to be seen. In conclusion, we have investigated the temperature and frequency dependence of the AC impedance properties of the superconducting samples and find out JJ decoupling to be one of the most important factors controlling the AC resistivity very much. As the AC impedance due to JJ decoupling is directly proportional to the applied AC frequency, at radio frequency and microwave frequency, it is expected to give bigger contribution.

**Acknowledgements:**

This work is supported by the Department of Science and Technology, University Grants Commission and the Council of Scientific and Industrial Research, Government of India.

**Figure Captions:**

1. SEM images of the surfaces of both the samples (sample 1 and sample 2). Both the samples show different microstructures due to the different preparation techniques.

2. Temperature dependence of the real part of complex resistivity for the sample labeled as sample 1 (YBCO sintered sample) and sample 2 (YBCO granular sample) at the frequency of 100 KHz. Solid curve is the Ambegaokar-Baratoff fit as explained in the text. We interpret the increase in resistivity with lowering temperature is due to the increase of JJ critical current of the junctions.

3. Temperature dependence of the real part resistivity for the sample 2 at different frequencies of 20, 40, 60, 80 and 100 KHz. Solid curves are the respective Ambegaokar-Baratoff fits $F(T)$ [(Eq. 2)] with delta (T) given by the BCS theory, as explained in the text.

4. Frequency (f) dependence of the in-phase part of resistivity for both the samples (sample 1 and sample 2) at $T = 30$ K and $f > 100$Hz. Solid curve is linear fit with different slope, as explained in the text.

5. Picture showing the position of JJ decoupling due the excessive current. $J_c$ is the critical current of the Josephson junction. The black rectangular box is the voltage-time area where the JJ exists. The arrow indicates the exact position at which JJ annihilation and formation occurs. Energy equal to $E_j$ is absorbed by the Josephson junction at the point "a" and the same amount of energy is released at the point "b". The energy released at the point "b" dissipates to the sample in the form of heat energy. This occurs twice at every cycle of AC current. One full cycle of AC gives two annihilations and two creations of the Josephson junction. The output voltage is the AC voltage given to the sample from the lock-in-amplifier.

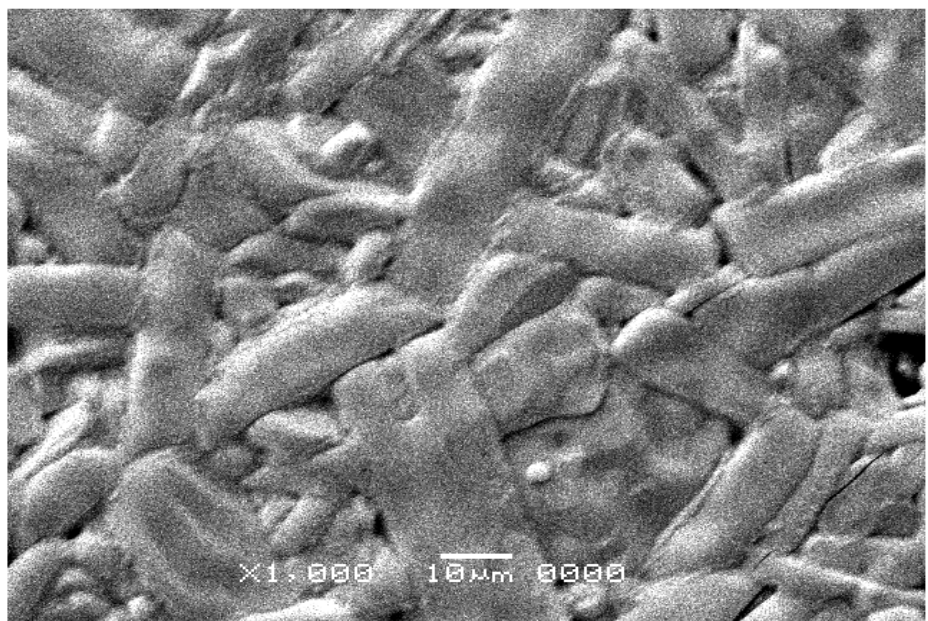

(Sample 1)

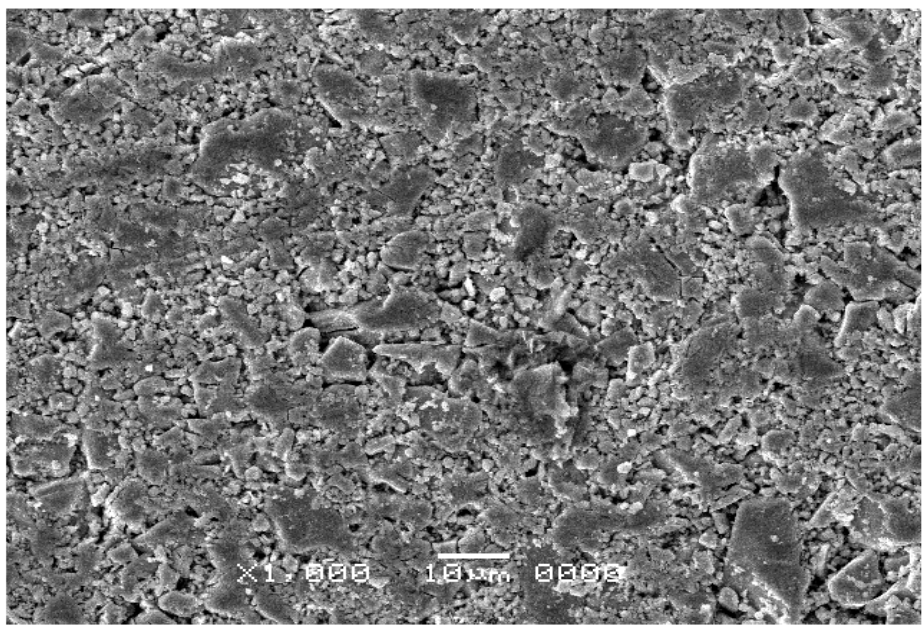

(Sample 2)

**FIG. 1**

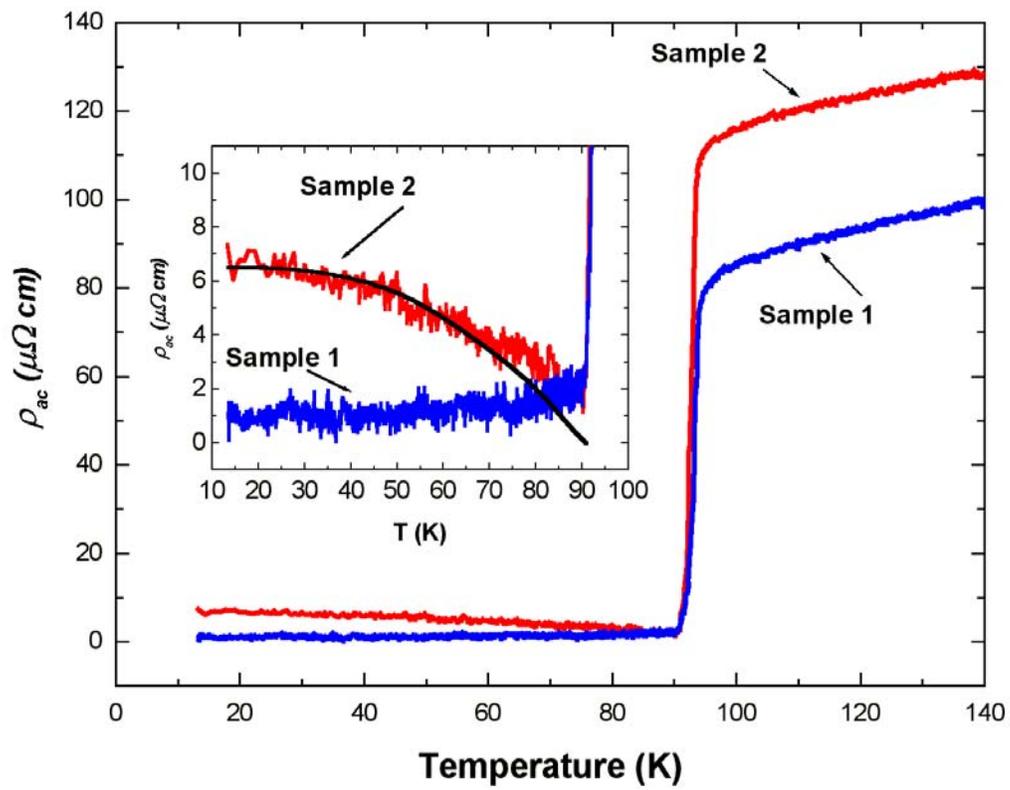

**FIG. 2.**

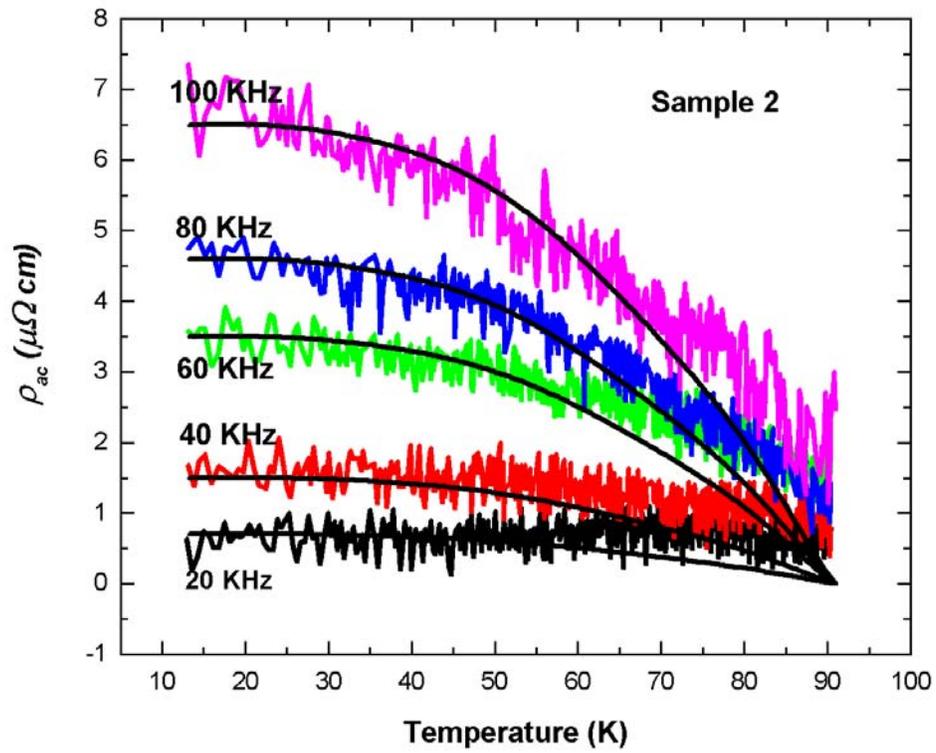

**FIG. 3.**

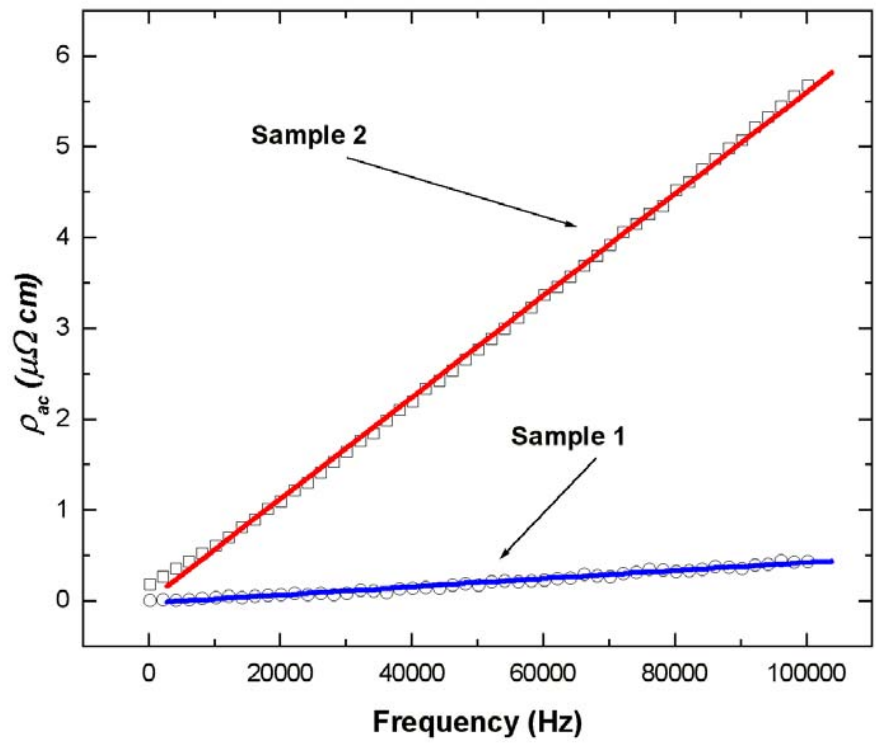

**FIG. 4.**

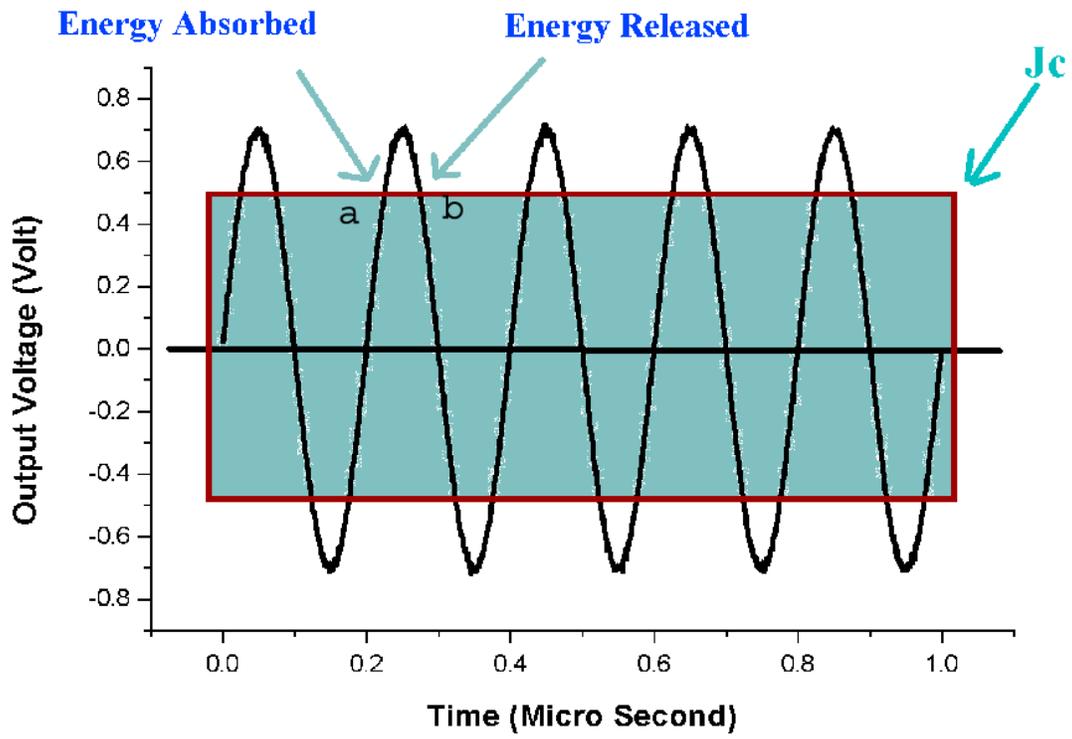

**FIG. 5.**